\documentstyle [12pt]{article}
\def\be{\begin{equation}}
\def\ee{\end{equation}}
\def\bea{\begin{eqnarray}}
\def\eea{\end{eqnarray}}

\begin{document}

\begin{titlepage}

\title{Integrable systems in spaces of arbitrary dimension}

\author{A.N. Leznov\\
{\it  Institute for High Energy Physics, 142284 Protvino,}\\{\it
Moscow Region,
Russia}
{\it and}\\ 
{\it  Bogoliubov Laboratory of Theoretical Physics, JINR,}\\
{\it 141980 Dubna, Moscow Region, Russia}}

\date{}

\maketitle

\begin{abstract}
The $2n$ dimensional manifold with two mutually commutative operators
of 
differentiation is introduced. Nontrivial multidimensional integrable
systems 
connected with arbitrary graded (semisimple) algebras are constructed. 
The general solution of them is presented in explicit form.
\end{abstract}
\end{titlepage}

\section{Introduction}

The success in the application of group-theoretical methods to the
theory of 
two-dimensional integrable systems \cite{1},\cite{2} is not
accidental. It 
is connected with the 
circumstance that the operations of multiplication of the group
elements from 
left and right are mutually commutative. It allows us  to associate
with 
two commutative differential operators $\frac{\partial}{\partial x},
\frac{\partial}{\partial y}$ of two dimensional space 
infinitesimal displacement generators dependent upon one parameter
from left and right  in such a way as to 
solve trivially the representation of zero curvature type.

The foundation of the whole construction is the group valued function:
\bea
K=M(x)N(y)\label{1}
\eea    
where the group valued elements $M,N$ are constructed by definite
simple rules,
containing only operations in one dimensional spaces  $(x,y)$
respectively.

Many different attempts have been made to generalize this construction
to the
multidimensional case by substituting instead of $(x,y)$ in (\ref{1})
some 
multidimensional functions $x=\phi(x_1,...x_n),y=\psi(y_1,...y_m)$ and
considering the consequences which follow from the associated
multidimensional
$L-A$ pair formalism. However no interesting nontrivial integrable
system has 
been discovered in this way. Moreover the solutions which arise in
such a 
construction (in the concrete examples  considered ) are always only
particular but 
not general. The  reader can find the corresponding literature in
\cite{RS}.

In the present paper we want exploit the following observation. It is
always 
possible to represent  general solutions of  integrable systems
following from 
(\ref{1})  in  local form. Or in other words to obtain the general
solutions of
such systems only the operation of differentiation is necessary (see
\cite{LS}). 
The operation of integration arises only on the middle step of
calculations and
may be eliminated from the final result. This means that if we  have a 
multidimensional space with two mutually commutative operators of 
differentiation $D_1,D_2$ we will be able to use the two dimensional 
construction to obtain nontrivial multidimensional integrable systems
together 
with their general solutions.

The aim of the present paper is in the explicit construction of
multidimensional
manifolds with such properties. The paper is organized in the
following way.
In section 2 we discuss  general properties of the operators of
differentiation
and obtain the equations they define. In section 3 we find the
solution of the
corresponding equations and in this realize the manifold with the 
necessary properties. Section 4 is devoted to consideration of
concrete 
examples of integrable systems on the multidimensional manifold
constructed 
above. 
Comments and perspectives for further investigation are collected in
section 5.

\section{The main equations defining the manifold} 

Suppose we have a $2n$ dimensional Euclidian space with a set of
independent 
coordinates $y_i,\bar y_i$. We assume the existence of two mutually
commutative
generators of differentiation $D,\bar D$, which satisfy the basic
relations:
\bea
D \bar y=0,\quad \bar D y=0\label{2}
\eea
which are reminiscent of (anti) holomorphic functions in the theory of
the 
complex variables. So $D$ is a linear combination of space derivatives
with 
respect to the unbarred coordinates, $\bar D$ is the same with respect
to the 
barred ones:
\bea
D=\frac{\partial}{\partial y_n}+\sum u^{\mu} \frac{\partial}{\partial
y_{\mu}},
\quad 
\bar D=\frac{\partial}{\partial \bar y_n}+\sum v^{\nu}
\frac{\partial}{\partial 
\bar y_{\nu}}.\label{3}
\eea 
We choose the coefficient functions of differentiation with respect to
$y_n,
\bar y_n$  to be equal to unity. This is an inessential restriction.

If we demand mutual commutativity of these generators $[D,\bar D]=0$,
then as a 
corollary of (\ref{3}) we obtain the system of equations, which the
coefficient
functions must satisfy: 
\be
Dv^{\nu}\equiv v^{\nu}_{y_n}+\sum u^{\mu} v^{\nu}_{y_{\mu}}=0,\quad 
\bar Du^{\mu}=u^{\mu}_{\bar y_n}+\sum v^{\nu} u^{\mu}_{\bar
y_{\nu}}=0.\label{4}
\ee
We will call (\ref{4}) the $uv$ system and the corresponding manifold
the $UV$ 
one. 

From (\ref{4}) it follows that  an arbitrary function $\bar f$ of
variables
$v,\bar y$ are annihilated by differentiation by $D$, all functions
$f(u,y)$ by  $\bar D$.
In this sense we will speak about holomorphic and antiholomrphic
functions. 

The following proposition arises:

Proposition 1.

Each function $\bar f$ annihilated by  the operator $D$  is
holomorphic; 
a function $f$ annihilated by the operator $\bar D$ is
antiholomorpthic.

Let us add  the equations $D \bar f=0$ and $\bar D f=0$ to the system
of $(n-1)$
equations (\ref{4}). Then the $n$ sets of variables $(1,u)$, and
$(1,v)$ 
respectively satisfy a linear system of $n$ algebraic equations the
matrix of 
which coincides with the Jacobian matrix (we consider the holomorphic
case): 
\be
J=\det_n \left|\begin{array}{cccc} v^1 & \dots & v^{n-1} & f \\
                    y_1 & \dots & y_{n-1} & y_n \end{array}\right|
\nonumber
\ee
which in the case of a non-zero solution of the linear system must
vanish. So Proposition 1 is proved.

As a corollary we obtain the following  

Proposition 2
\be
\bar Dv^{\nu}=v^{\nu}_{\bar y_n}+\sum v^{\mu} v^{\nu}_{\bar y_{\mu}}=
V^{\nu}(v;\bar y),\quad 
Du^{\mu}=u^{\mu}_{y_n}+\sum u^{\nu}
u^{\mu}_{y_{\nu}}=U^{\mu}(u;y)\label{BA}
\ee
Indeed operators $D,\bar D$ are commutative and so $\bar Dv$ is
solution of the 
same equation as $v$ satisfies. But each solution of third equation is
a
holomorphic function, which proves  proposition 2.

If we consider functions $U,V$ as given, then all operations of 
differentiation applied to the functions $\Phi(u,v;y,\bar y)$ are well
defined.

As was mentioned in the introduction only these operations arise in
the theory
of integrable systems constructed in the manner of (\ref{1}). Thus for
a
realization of the proposed program it is necessary to solve the
system of
equations (\ref{3}),(\ref{BA}) with the given holomorphic and
antiholomorphic
$U,V$ functions.

\section{General solution of $uv$ system and realisation of the
multidimensional
$UV$ manifold}  

Consider the system of equations defining implicitly $(n-1)$ unknown 
functions $(\phi)$ in $(2n)$ dimensional space $(y,\bar y)$:
\bea 
Q^{\nu}(\phi;y)=P^{\nu}(\phi;\bar y)\label{D}
\eea
with the convention that all Greek indices take values between $1$ and
$(n-1)$.
The number of equations in (\ref{D}) coincides with the number of
unknown 
functions $\phi^{\alpha}$. Each arbitrary function $Q,P$ depends on
$(2n-1)$
coordinates.

With the help of the usual rules of differentiation of  implicit
functions
we find from (\ref{D}):
\bea
\phi_y=(P_{\phi}-Q_{\phi})^{-1} Q_y,\quad \phi_{\bar
y}=-(P_{\phi}-Q_{\phi})^
{-1}P_{\bar y}\label{DD}
\eea
Let us assume, that between $n$ derivatives with respect to barred and
unbarred
variables the following linear dependence takes place:
$$
\sum^n_1 c_i \phi^{\alpha}_{y_i}=0,\quad \sum^n_1 d_i
\phi^{\alpha}_{\bar y_i}=0
$$
and analyse the corollaries following from these facts.

Assuming that $c_n\neq 0,d_n\neq 0$, dividing them into each 
equation of the left and right sets respectively and introducing the
notation 
$v^{\alpha}={c_{\alpha}\over c_n},u^{\alpha}={d_{\alpha}\over d_n}$,
we rewrite 
the last system in the form:
\bea
\phi^{\alpha}_{y_n}+\sum^{n-1}_1 v^{\nu}
\phi^{\alpha}_{y_{\nu}}=0,\quad
\phi^{\alpha}_{\bar y_n}+\sum^{n-1}_1 u^{\nu} \phi^{\alpha}_{\bar
y_{\nu}}=0
\label{MS} 
\eea
Substituting values of the derivatives from (\ref{DD}) and multiplying 
result by the matrix $(P_{\phi}-Q_{\phi})$ on the left we obtain:
\bea
Q^{\alpha}_{y_n}+\sum^{n-1}_1 v^{\nu} Q^{\alpha}_{y_{\nu}}=0,\quad
P^{\alpha}_{\bar y_n}+\sum^{n-1}_1 u^{\nu} P^{\alpha}_{\bar
y_{\nu}}=0\label{D1}
\eea
From the last equations it immediately follow that:
\bea
v^{\nu}=-(Q_y)^{-1} Q_{y_n},\quad u^{\nu}=-(P_{\bar y})^{-1} P_{\bar
y_n}
\label{UV}
\eea
We see that if we augment the initial system (\ref{D}), by $(n-1)$
vector 
functions $(u,v)$ defined by (\ref{UV}) then the differential
operators $D
\bar D$ defined by (\ref{3}) in connection with (\ref{MS}) annihilate
every 
$\phi$ and therefore the functions $Q,P$ :
\bea
D \phi=\bar D \phi=D Q=D P=\bar D Q=\bar D P=0 \label{VIC}
\eea
This means that $D\bar f(\phi,\bar y)=\bar D f(\phi, y)=0$. And as a
direct
corollary of this fact $Dv=\bar D u=0$ and so the generators $D,\bar
D$ 
constructed in this way are mutually commutative. 

Thus we have found the general solution of the $uv$ system and in such
a way 
realise the manifold with the properties postulated in the previous
section. 

It is possible to say (the solution of the $uv$ system is  general)
that
this realisation is unique up to possible similarity transformations.

We present below the result of calculations of the functions $U,V$
using the 
definition of $u,v$ functions (\ref{UV}):
$$
U=D u=-Q_y^{-1}(D Q_{y_n}+\sum u^{\alpha} D Q_{y_{\alpha}}) ,\quad
V=\bar D v=-P_{\bar y}^{-1}(\bar D P_{\bar y_n}+\sum ^{\alpha} \bar D
Q_{\bar 
y_{\alpha}})
$$

\section{Examples}

Below we would like to consider only two examples clarifying the
situation. But
really the same is true with respect to each two dimensional system
integrable 
with the help of the formalism of graded algebras \cite{1},\cite{2}.

We want to emphasize that in all cases (particulary those considered
below) the 
$UV$ manifold by itself is defined by $2(n-1)$ arbitrary functions
$Q,P$ each of 
one of which  depends upon $2n-1$ independent arguments. All these
functions
occur as coefficient functions (via operators of differentiation
$D,\bar D$) in equations of multidimensional integrable systems.
Naturally the 
general solution  depends upon them. Apart from these functions the
general 
solution of an integrable system depends upon additional arbitrary
functions 
the number of which and their functional dependence has to be
sufficient for 
the statement of Cauchy or Gursat initial data problems. 

\subsection{Multidimensional Liouville equation}

By this term we understand the equation:
$$
D\bar D \phi=\exp 2\phi
$$
By direct calculation one can become convinced that its general
solution has 
the form:
$$
\exp-\phi=(c(u;y)+\bar c(v;\bar y) (D c)^{{1\over 2}} (\bar D \bar c)^
{{1\over 2}}
$$ 
For instance in the four-dimensional case $(y_1,y_2;\bar y_1,\bar
y_2)$:
$$
D\bar D=\bar D D=\frac{\partial^2}{\partial y_2 \partial \bar y_2}+
v\frac{\partial^2}{\partial y_2 \partial \bar y_1}+u\frac{\partial^2}
{\partial \bar y_2 \partial y_1}+uv\frac{\partial^2}{\partial y_1
\partial 
\bar y_2}
$$ 
and 
$$
u=-{Q_{y_2}\over Q_{y_1}},\quad u=-{P_{\bar y_2}\over P_{\bar
y_1}},\quad
Q(\phi;y_1,y_2)=P(\phi;\bar y_1,\bar y_2)
$$

\subsection{Multidimensional Toda system}

The equations of the Toda lattice in two dimensions have the form (one
of the 
many possible ones):
$$
x^i_{z,\bar z}=\exp (-x^{i-1}+2x^i-x^{i+1}),\quad 1\leq i \leq n,\quad
x_{-1}=x_{n+1}=0
$$
The general solution may be represented as:
$$
\exp-x^i=\det_i \{ V^0 \},\quad \exp-x_0=V^0
$$
where matrix elements of the determinant matix $V^0$ are as follows:
$$
V^0_{i,j}=\frac{\partial^{i+j-2}}{\partial z^{i-1}\partial \bar
z^{j-1}} V^0
$$
and the single function $V^0$ has the form:
$$
V^0={W_n(\theta,z) W_n(\bar \theta,\bar z)\over (1+\sum_1^n
\theta_i\bar 
\theta_i)}
$$ 
$W_n(\phi,x)$ is the determinant of the  Wronskian matrix with
elements:
$$
W_{i,j}=\frac{\partial^{i-1}\phi^j}{\partial x^{j-1}} 
$$

To obtain the explicit general solution of multidimensional Toda
lattice system:
$$
\bar D D=D\bar D x^i=\exp (-x^{i-1}+2x^i-x^{i+1})
$$
only the following changes in all the corresponding formulae above are 
necessary:
$$
\frac{\partial}{\partial z}\to D,\quad \frac{\partial}{\partial \bar
z}\to \bar
D
$$
and consider arbitrary functions $\theta,\bar \theta$ as arbitrary
holomorphic,
antiholomorphic functions on the manifold $UV$ of the corresponding
dimension.

\section{Outlook}

The main results of the present paper are in the general solution of
the $uv$ 
system (\ref{4}), a realisation of a multidimensional manifold with
two 
commutative operators of differentiation (section 3) and 
the integrable systems in the spaces of arbitrary dimensions
constructed in 
this framework (section 4).

It turns out that with the help of $UV$ manifold it is possible to
find
general solution of such interesting from the point of viw physical
applications
systems as homogeneous Complex Monge-Ampher and Bateman equations in
the spaces of
arbitrary dimensions \cite{combat}. This was achieved by reducing of
the
definite kind the general solution of $uv$ system on subclasses in
which solution of it is functionally depends only on two arbitrary
functions of
the necessary number of independent arguments.

It is obviousl that formalism of evolution type systems (integrable
with
the help of the old inverse scattering method) remains 
without any changings in the spaces of arbitrary dimension. The
exactly
integrable systems considered above are the symmetries of evolution
type
equations. The knowledge of the general solution of the first allow
represent
in explicit form multisoliton solutions of the last.

We should like to finish this outlook with some speculative comments
about the 
possible application of the  proposed construction to the problems of
the 
physics.

If the manifold constructed would have any relation to the real four
dimensional
world, then something similar to Einstein's General Relativity would
occur. 
Indeed, in both cases the general (geometrical) properties of the
world be 
determined by some fundamental physical objects, the metrical tensor
$g_{ij}$ in 
the case of General Relativity, Einstein's equations for which takes
into account 
the presence of all forms of matter and the equations $uv$ describing
the 
manifold $UV$ in the case considered above (which of course must be
modified to
take into account all other physical fields). To be optimistic, it may
happen 
that equations of General Relativity have a solution on the manifold
of the 
kind described above or something similar to it. Of course all this is
only an 
attractive speculation and only a deeper investigation of the problem
may 
clarify the situation.

\section*{ Acknowledgements.}

The author gratefully thanks D.B.Fairlie during the common work with
whom
on the problems of the Monge-Amph\'ere and the Bateman equation the
idea of the manifold $UV$
was born, for  discussions in the process of working on this paper and 
important comments.

The author is indebted to the Center for Research on Engenering
and Applied Sciences (UAEM, Morelos, Mexico) for its hospitality and
Russian Foundation of Fundamental Researches (RFFI) GRANT N
98-01-00330 for 
partial support.

\end{document}